\documentclass[osajnl,preprint,showpacs,superscriptaddress,12pt]{revtex4-1} 
\usepackage{amsmath,amssymb,graphicx}
\usepackage{blkarray}
\begin{document}

\title{Time-resolved two-photon excitation of long-lifetime polaritons}

\author{Chitra Gautham}\email{Corresponding author: chg48@pitt.edu}
\author{Mark Steger}
\author{David Snoke}
\affiliation{Department of Physics and Astronomy, University of Pittsburgh, Pittsburgh, PA 15260, USA}

\author{Ken West}
\author{Loren Pfeiffer}
\affiliation{Department of Electrical Engineering, Princeton University, Princeton, NJ 08544, USA}

\begin{abstract}Recent studies of two-photon excitation of exciton-polaritons in microcavities have considered the possibility of an allowed absorption process into the $2p$-state of  the excitons which participate in the polariton effect. Here we report time-resolved measurements of two-photon excitation directly into the lower polariton states invoking the $1s$ state of the excitons. Although this process is forbidden by symmetry for light at normal incidence, it is allowed at non-zero angle of incidence due to state mixing.
\end{abstract}

\maketitle 

The exciton-polariton is a quantum superposition of light and matter which has been studied extensively for its bosonic properties. The canonical system consists of quantum well (QW) excitons embedded in a two-dimensional microcavity (for reviews see, e.g., Refs. \cite{Deng,Carusotto,Kav,rev1}).  At low temperatures, this system exhibits Bose-Einstein condensation \cite{Kaz,Balili,Yamo}, superfluidity \cite{Amo,super1,super2}, and quantized vortices \cite{Gang,DevVort,Vert1,Vert2,Vert3} and may have applications as a low-threshold coherent light source and highly nonlinear optical system. Polaritons are metastable particles, as they can leak through the mirrors into external photons. 

\begin{figure}
\centerline{\includegraphics[width=0.7\columnwidth]{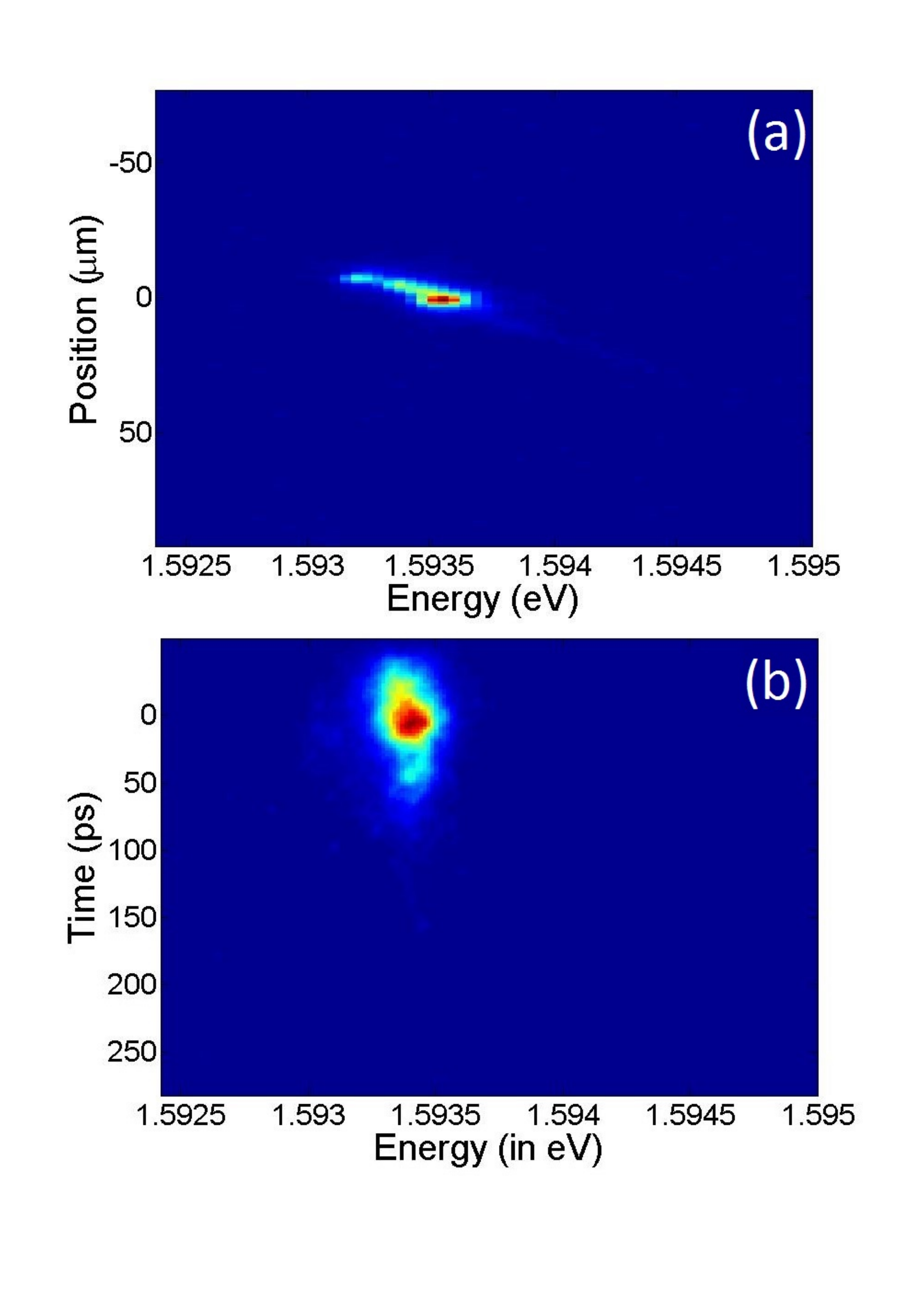}}
\caption{(a) Time-averaged spectrum while exciting with half the energy of the LP with a laser focused at $x=0\mu m$. There is a spatial gradient to the polariton energy because there is a wedge in the cavity thickness across the sample. (b) Time-resolved spectrum at the generation spot while exciting with half the energy of the LP. There is a fast rise time of the emission comparable to our time resolution which indicates there is direct excitation of the the LP.}
\end{figure}

Recently, two-photon excitation of exciton-polaritons has gained attention \cite{kavokin, Sch, Bloch} for its possible application in polariton lasers \cite{kavokin, ref18, ref19}. In general, two-photon excitation of polaritons provides another way of using the super-nonlinearities of the polaritons for optical modulation schemes. Refs.~\cite{kavokin, Bloch} proposed a mechanism via absorption into the 2p state while ref.~\cite{Sch} investigated this claim using a narrowband source and cast doubt on that mechanism because they saw no strong two-photon absorption at the energy of the 2p state. Since there exist other mechanisms that could permit the conversion of dark excitons into the lower polaritons, in this paper we perform time-resolved measurements to investigate this and show successful direct excitation of exciton-polaritons by two-photon absorption \cite{poster}. We show that this is possible with a beam incident with a finite in-plane momentum due to ``bright" state/``dark" state mixing and study the polarization dependence of this absorption both theoritically as well as experimentally.

In the GaAs-based structures we use, the lowest quantum well exciton states consist of the two $J=1$ ``bright" states and two $J=2$ ``dark" states. These states are made of the conduction electrons with spin =$\pm1/2$ and the heavy holes with angular momentum $\pm3/2$. In our samples with narrow (7 nm) quantum wells, the light hole exciton states are about 30 meV higher than the heavy hole states, which is greater than the upper polariton-lower polariton splitting and greater than the spectral width of the lasers we use, so the light hole states are unlikely to be excited in resonant excitation of the heavy hole states.  The polaritons are formed by a coupling of the cavity photon to the $J=1$ ``bright" heavy hole exciton state.  The coupling of the photon and exciton states leads to two new sets of states, the upper polaritons' and lower polaritons,' split by about 12 meV.  The $J=2$ ``dark" exciton states do not couple to photons to make polaritons.

Our exciton-polariton samples are made up of GaAs quantum wells with AlAs barriers, in three sets of four placed at the antinodes of the microcavity made of two Bragg mirrors, which made of AlAs and Al$_x$Ga$_{1-x}$ repeating layers. The details of these long lifetime samples is given is Ref.~\cite{prx} and the measurement of the lifetime ($180\pm10$ ~ps) is given in Ref.~\cite{life}.  This lifetime corresponds to a quality factor of over 300,000, compared to a quality factor less than 10,000 for the samples used in Refs.~\cite{Sch, Bloch}. The long lifetime of the samples allowed us to study ballistic propogation of the population injected by two-photon excitation. The ballistic propogation was good evidence that the injection was directly into polariton states rather than higher energy states that would have to thermalize/scatter into the LP states.

These samples were mounted in a cryostat and held at a fixed temperature in the range of 4-8 K. Two-photon excitation of the samples was done using a Coherent optical parametric amplifier (OPA) system consisting of a femtosecond pulsed Ti-Sapphire laser,  a regenerative amplifier with a repetition rate of 250 kHz, and an OPA pumped with tunable output wavelength. Since the energy of the lower polariton in our samples was about 1.593 eV, we tuned the OPA to give a beam with half that energy (0.7965 eV). This output beam had a spectral full width at half maximum (FWHM) of 15~meV.

\begin{figure}
\centerline{\includegraphics[width=0.8\columnwidth]{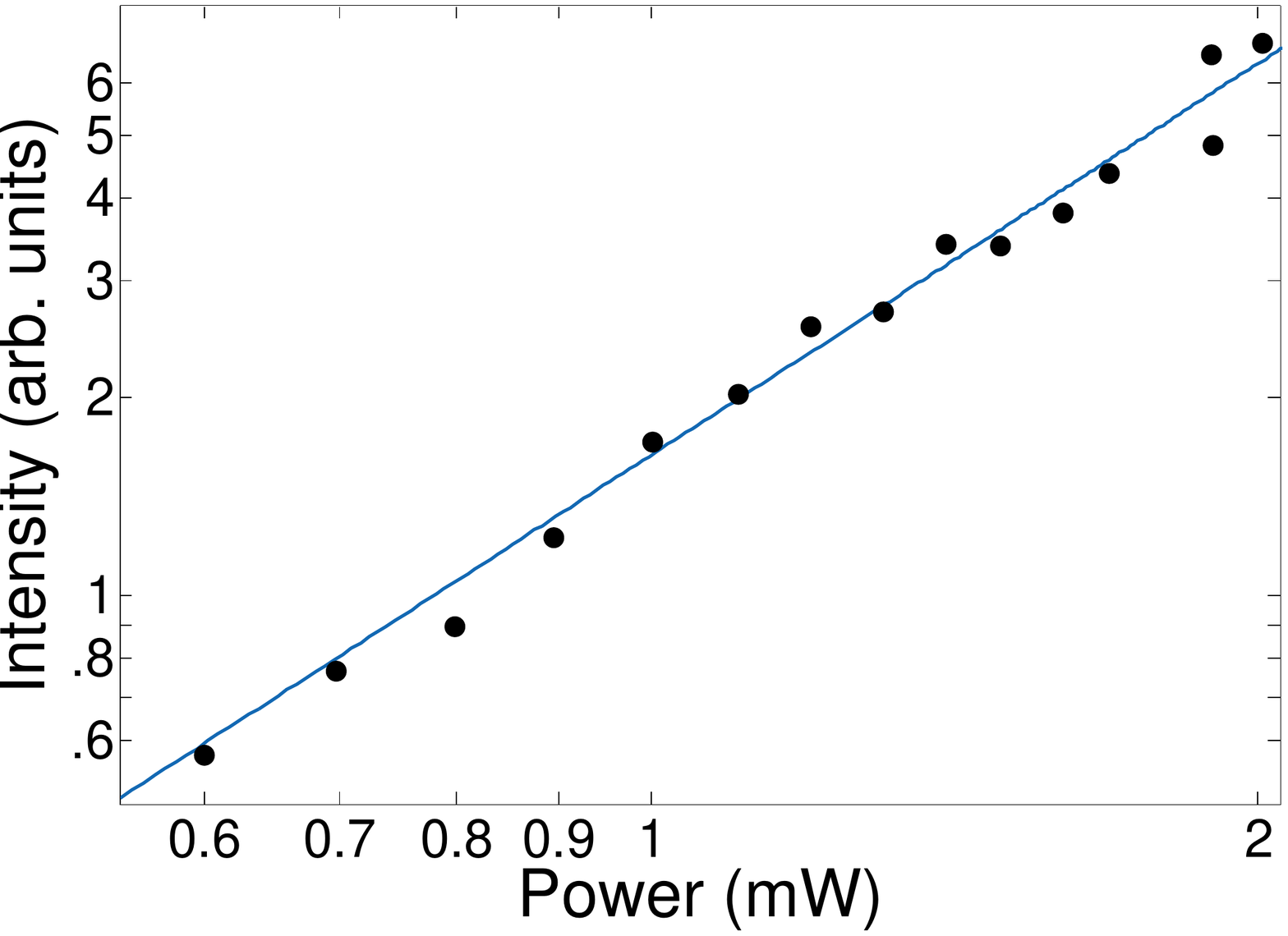}}
\caption{Power dependence of the polariton emission intensity at the creation spot. Solid line: fit to the square of the pump power, indicating two-photon absorption.}
\end{figure}

We used a dichroic mirror and a 1000~nm long pass filter in the path of our pump beam in order to remove leaked signal coming from the regenerative amplifier. Also, since the photon energy of the regenerative amplifier beam is much lower than the energy of the polariton emission, it was not detected by our spectrally-resolved detection system. The emission signal from the polaritons was spectrally resolved using a 0.25-m spectrometer and time-resolved using a Hamamatsu streak camera. The time-averaged signal was simultaneously viewed on a Princeton CCD camera. Figure 1(a) shows a typical spatially resolved, time-averaged spectrum. The energy of the emission from the polaritons varies across the sample because there is a wedge in the cavity thickness, giving a spatial gradient to the photon energy. Figure 1(b) shows a typical time-resolved spectrum. As seen in this figure (as well as Figure 3(b)), there is a fast rise time of the emission, comparable to our time resolution. The fast fall time of the emission is actually an artifact due to the motion of the polaritons in the cavity gradient. Instead of remaining at the spot where they were generated, the polaritons accelerate in the direction toward lower cavity photon energy. This leads to two effects that suppress their collection by our detection system. First, as they move, they can move out of the spatial field of view of the lens collecting the emission. Second, as they accelerate to higher momentum in the plane (corresponding to higher in-plane wave number $k_{\|}$), their photon emission occurs at higher angle, and therefore will not be collected by a low numerical-aperture system.

\begin{figure}
\centerline{\includegraphics[width=1.4\columnwidth]{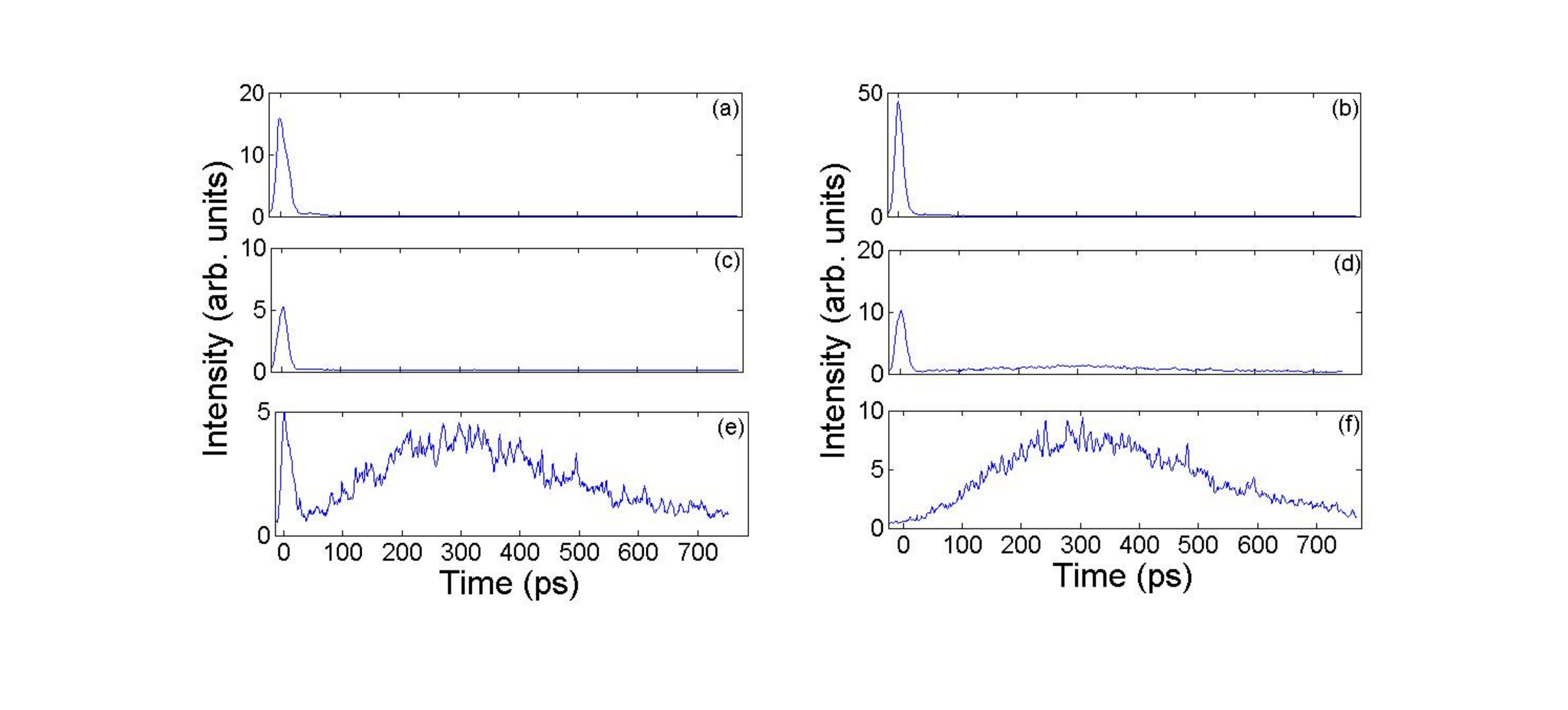}}
\caption{Intensity vs.~time for different pump wavelengths of (a)1565nm (b)1555 nm (c)1540 nm (d)1530 nm (e)1525 nm and (f) 1515 nm. A fast initial peak appears when the median pump photon energy is at half the lower polariton energy ($\lambda = 1555$~nm). A later population dominates for higher pump energies.}
\end{figure}

In order to ensure that we were observing two-photon excitation and not a higher-order excitation or single-photon excitation due to leaked photons in the pump beam, we did a power series measurement by varying the pump power and measured the time-averaged intensity. As seen in Figure 2, the good fit to $I \propto P^2$ power law confirms that we have observed two-photon excitation.

Figure 3 shows the results of the time-resolved measurements for various pump wavelengths. The wavelength of 1555 nm corresponds to the resonant condition of the pump photon energy exactly half the lower polariton energy. 
When exciting with exactly half the resonant energy (Figure 3(b)), we see a short (16 ps) peak. However, as we increased the pump photon energy, we see the initial peak disappear and a signal with a long rise time take its place.

Figure 4 shows the polariton intensity at constant pump power as the pump wavelength is varied. When we plot only the intensity of the initial peak, as shown in Figure 4(a), we see that the intensity is maximum when the pump photon energy is half the LP energy  and disappears at a higher pump photon energy. The FWHM of this peak is 15~meV, which is the same as the pump laser spectral FWHM.  If we plot the total intensity, as shown in Figure 4(b), we see that the intensity increases with increasing pump photon energy. We also observe a small peak while pumping at an energy corresponding to half the energy of the upper polariton energy (0.807 eV) in both cases.

\begin{figure}
\centerline{\includegraphics[width=0.861\columnwidth]{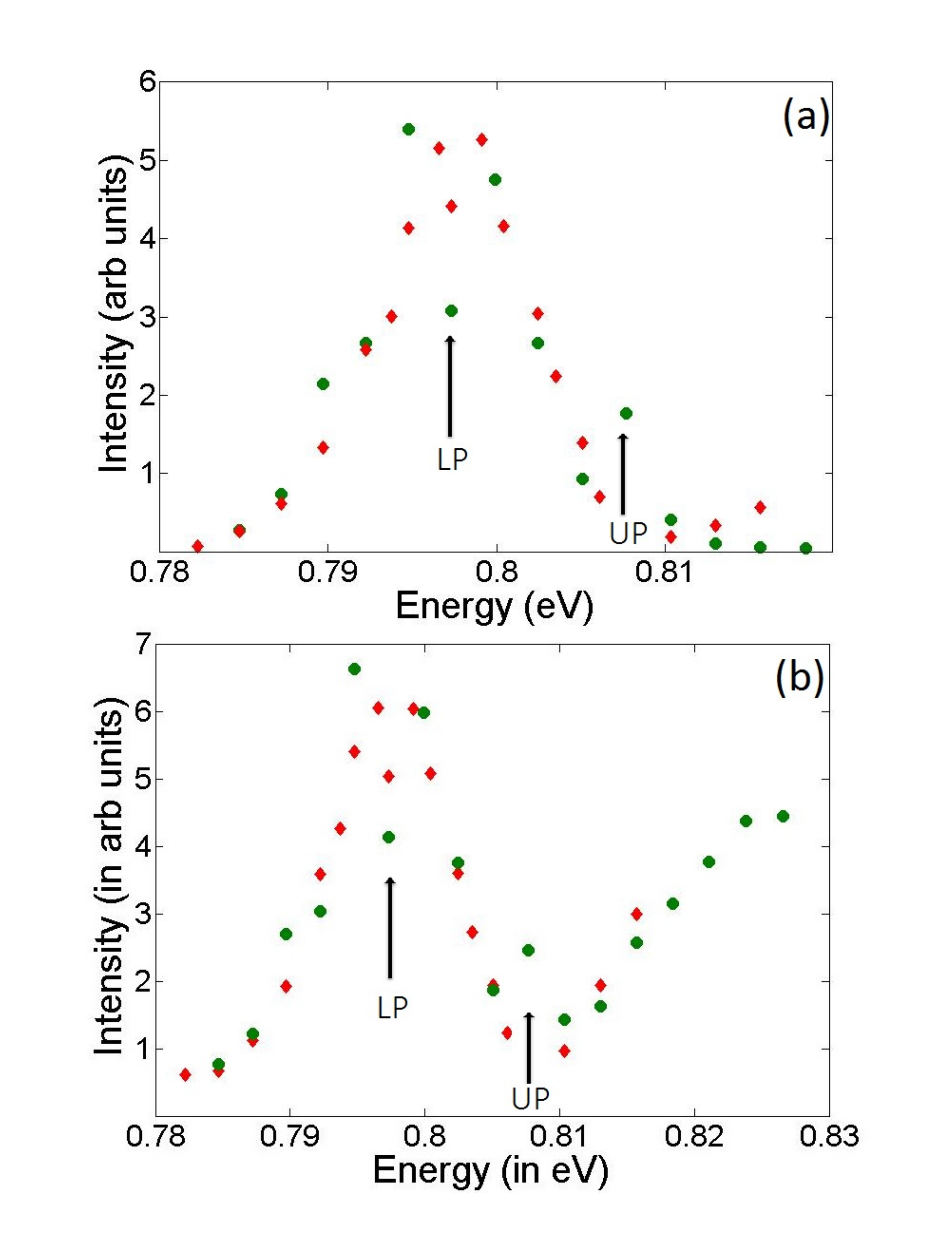}}
\caption{Polariton emission intensity vs.~pump photon energy from the time-resolved data. (a) Initial peak intensity showing maximum absorption at the LP energy with a slight peak at the UP energy. (b) Total integrated intensity showing an increase in intensity as the energy is increased. The circles and diamonds represent two data sets viewed on two different days.}
\end{figure}

For each wavelength, we measured the power dependence for the initial fast-risetime peak and the slow-risetime signal separately. Both signals had intensity proportional to the square of the pump power, indicating that both cases correspond to two-photon absorption. 

In order to understand the slow-risetime signal, we measured the signal as a function of temperature. As seen in Figure 5, when the temperature is lower, the slow-risetime signal has greater relative weight. This is consistent with higher energy states cooling down into the ground state of the lower polariton. At a higher temperature, these states will be scattered to higher $k$-states, while at a lower temperature, they can cool down to states near $k_{\|}=0$. Since $k=0$ corresponds to emission normal to the cavity, and our detection system has low numerical aperture, we observe only states with $k_{\|} \simeq 0$ in our experiment. The short initial peak shows no change in intensity in the temperature range we studied (2.5 K to 10 K).

The picture thus arises that polaritons are created by two different mechanisms. One mechanism is direct two-photon creation of polaritons, which occurs most efficiently when the pump laser photon energy is at exactly half the lower polariton energy. The second process is two-photon absorption into excitons in higher energy states, which then relax down into the lower polariton states with a time constant of several hundred picoseconds. These higher-energy states may be either ``dark" ($J=2$) exciton states or $2p$ states of the $J=1$ excitons. 

In order to explain the luminescence from two-photon absorption, other works~\cite{kavokin,Bloch} considered a mechanism based on absoprtion into the 2p-state of the exciton, which then relaxes into the lower polariton by emitting a terahertz photon, while ref.~\cite{Sch} has found no evidence of direct two-photon pumping into the 2-p state. Time resolving the luminescence (Figure 3 and 4) leads us to believe that we are directly exciting the polariton states.  Although two-photon excitation of the $J=1$ states is forbidden by symmetry at $k_{\|}=0$ (normal incidence), away from $k_{\|}=0$, mixing of the $J=1$ and $J=2$ states occurs.

The coupling between the ``dark" and ``bright" exciton states can be derived from the Luttinger-Kohn (L-K) Hamiltonian \cite{Chuang,Jeff}:
 
\begin{widetext}
$H^{LK}|u_{k}\rangle  =
\left( \begin{array}{cccccc}
  P+Q &  -S &  R &  0 &  -S/\sqrt{2} &  \sqrt{2}R \\
 -S^{*} &  P-Q &  0 &  R & -\sqrt{2}Q &  \sqrt{3/2}S \\
 R & 0 & P-Q & S & \sqrt{3/2}S & \sqrt{2}Q \\
 0 & R^{*} & S^{*} & P+Q & -\sqrt{2}R & -S^{*}/\sqrt{2} \\
 -S^{*}/\sqrt{2} & -\sqrt{2}Q^{*} & \sqrt{3/2}S & -\sqrt{2}R & P+\delta & 0\\
 \sqrt{2}R^{*} & \sqrt{3/2}S^{*} & \sqrt{2}Q^{*} & -S/\sqrt{2} & 0 & P+\delta\\ 
 \end{array} \right)$ 
{\large $
\begin{blockarray}{cc}
 J,m_j &  \\
\begin{block}{(c)c}
|\frac{3}{2},\frac{3}{2}\rangle &\\
|\frac{3}{2},\frac{1}{2}\rangle &\\
|\frac{3}{2},-\frac{1}{2}\rangle &\\
|\frac{3}{2},-\frac{3}{2}\rangle &\\
|\frac{1}{2},\frac{1}{2}\rangle &\\
|\frac{1}{2},-\frac{1}{2}\rangle &\\
\end{block}
\end{blockarray}$

 

}
\end{widetext}

where
 \begin{eqnarray}
P &=& \frac{\hbar\gamma_{1}}{2m_{0}}(k_{x}^{2}+k_{y}^{2}+k_{z}^{2}) \nonumber\\
Q &=& \frac{\hbar\gamma_{2}}{2m_{0}}(k_{x}^{2}+k_{y}^{2}-2k_{z}^{2}) \nonumber\\
R &=& \frac{\hbar}{2m_{0}}(-\sqrt{3}\gamma_{2}(k_{x}^{2}-k_{y}^{2})+i2\sqrt{3}\gamma_{3}k_{x}k_{y}) \nonumber\\
S &=& \frac{\hbar\gamma_{3}}{2m_{0}}\sqrt{3}(k_{x}-ik_{y})k_{z}.
\end{eqnarray}

Because $\delta$ is about 100 meV, we can ignore the split-off holes. We therefore restrict our attention to the $4\times 4$ submatrix the light-hole heavy-hole states.   The ``bright" ($J=1$) excitons for the heavy holes have conduction-band electrons with spin in the direction opposite to the hole angular momentum, i.e. $|\frac{3}{2},\pm\frac{3}{2};\mp\frac{1}{2}\rangle$, where the first label gives the value of $J$ for the hole, the second label gives the value of $m_j$ of the hole, and the third, the spin of the electron. ``Dark" heavy-hole ($J=2$) excitons correspond to electron spin in the same direction as the hole angular momentum, i.e., $|\frac{3}{2},\pm\frac{3}{2};\pm\frac{1}{2}\rangle$.

\begin{figure}
\centerline{\includegraphics[width=0.9\columnwidth]{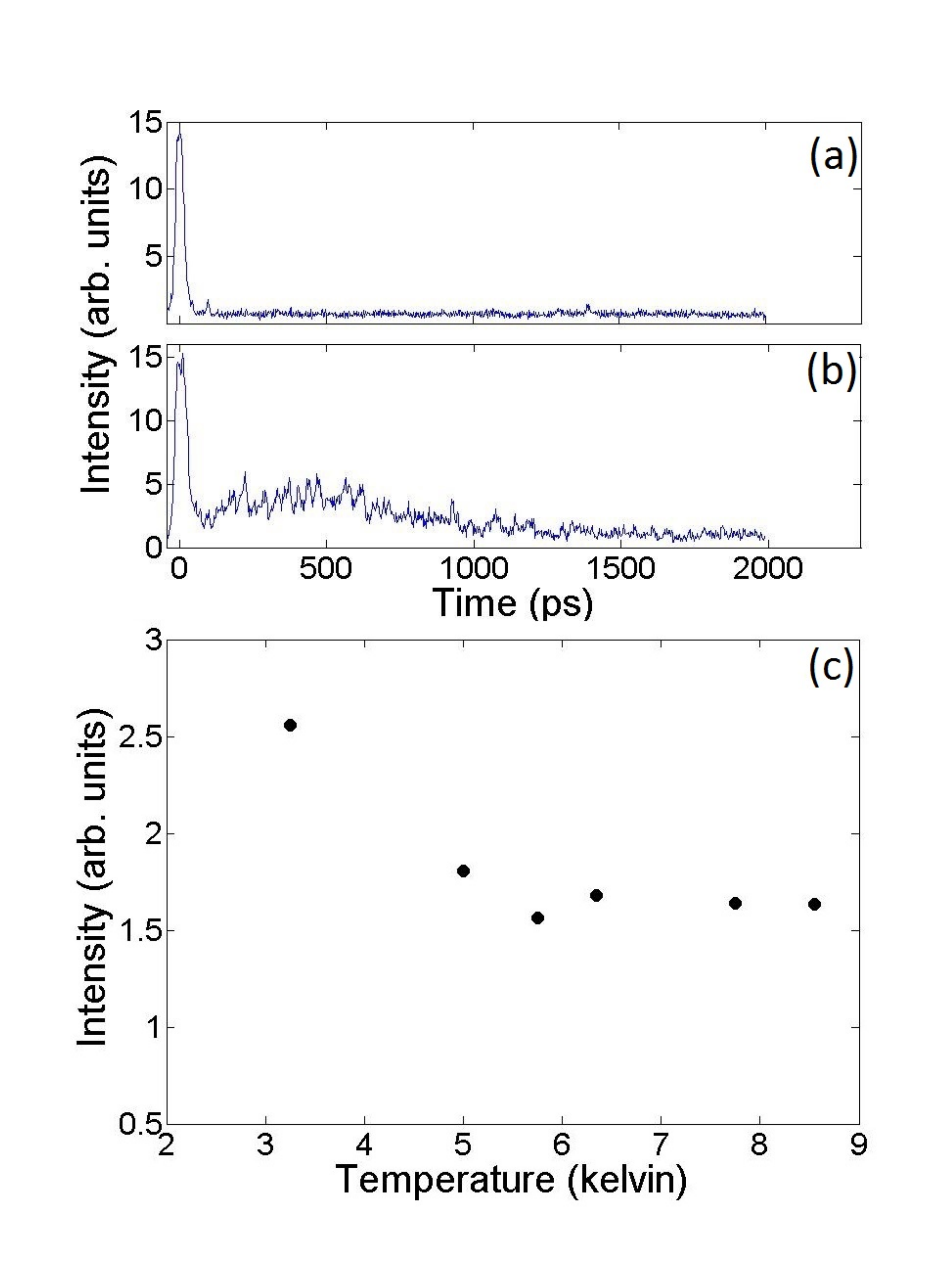}}
\caption{Time vs intensity at (a) 8.3K and (b) 2.5K. As the temperature is increased, the intensity of the latter peak decreases. (c) Summary of the late-time intensity data as a function of $T$.}
\end{figure}

Our pump photons are linearly polarized. Since linear polarization can be viewed as the superposition of two opposite circular polarizations, two photons from the pump beam will couple to a net $J=0$ state of the excitons.  Such a state exists for the light holes, corresponding to the two states $|\frac{3}{2},\frac{1}{2};-\frac{1}{2}\rangle$ and $|\frac{3}{2},-\frac{1}{2};\frac{1}{2}\rangle$. The ``bright" state polaritons are nominally defined as $\alpha|\frac{3}{2},-\frac{1}{2}\rangle+\beta|1\rangle$ and $\alpha'|-\frac{3}{2},\frac{1}{2}\rangle +\beta'|-1\rangle$, where $|\pm 1\rangle$ are the cavity photon states with $J=1$.  
The $J=0$ light hole states couple to the $J=1$ heavy hole states through the L-K Hamiltonian when $S \ne 0$. The $S$ term increases linearly with $k_{x}$ and $k_{y}$ and is zero for $k_{x}=k_{y}=0$. The value of $k_z$ is determined by the quantum well confinement. 

We thus see that the lower polaritons are not purely made from heavy hole excitons for finite in-plane $k_{\|}$; the excitonic part of the polariton includes a ``dark" exciton fraction. The ``dark" exciton fraction will slightly reduce the polaritonic coupling to the cavity photons but will not lead to drastic changes of the polariton behavior.
Diagonalizing the L-K Hamiltonian, setting $R = 0$ and  $P+Q = E_{hh}$ and $P-Q = E_{lh}$, and $\Delta = E_{lh}- E_{hh}$, which as mentioned above is about 30 meV, and $S \ll \Delta$, the hole eigenstates are
\begin{eqnarray}
&|\textstyle\frac{3}{2},\frac{3}{2}\rangle + \displaystyle \frac{S}{\Delta } |\textstyle \frac{3}{2},\frac{1}{2}\rangle \nonumber\\
&\displaystyle- \frac{S}{\Delta }|\textstyle \frac{3}{2},\frac{3}{2}\rangle + |\frac{3}{2},\frac{1}{2}\rangle \nonumber\\
&|\textstyle\frac{3}{2},-\frac{3}{2}\rangle + \displaystyle \frac{S}{\Delta } |\textstyle \frac{3}{2},-\frac{1}{2}\rangle \\
&\displaystyle- \frac{S}{\Delta }|\textstyle \frac{3}{2},-\frac{3}{2}\rangle + |\frac{3}{2},-\frac{1}{2}\rangle \nonumber . 
\end{eqnarray}

Although in these experiments we excited the sample at normal incidence, we used a focusing lens which introduced a finite range of angles of incidence, and therefore finite $k_{x}$ and $k_{y}$, which caused direct excitation of the LP through coupling with the ``dark" state excitons. The above interpretation implies that an increase in $k_{x}$ and $k_{y}$ should increase the absorption of the two photon absorption. 

We further note that the eigenstates of the light holes are given by \cite{Chuang}

\begin{equation}
\begin{split}
|\textstyle \frac{3}{2},\frac{1}{2}\rangle = \frac{-1}{\sqrt 6}|(\cos\theta \cos\phi - i \sin\phi)\hat{x} + (\cos\theta \sin\phi + i \cos\phi)\hat{y}\\ -\sin\theta \hat{z}\rangle|\downarrow\rangle \\+\textstyle\sqrt{\frac{2}{3}} |\sin\theta \cos\phi \hat{x} + \sin\theta \sin\phi \hat{y}+\cos\theta \hat{z}\rangle|\uparrow\rangle 
\end{split}
\end{equation}
\begin{equation}
\begin{split}
|\textstyle \frac{3}{2},-\frac{1}{2}\rangle = \frac{-1}{\sqrt 6}|(\cos\theta \cos\phi+i \sin\phi)\hat{x} + (\cos\theta \sin\phi - i \cos\phi)\hat{y}\\-\sin\theta \hat{z}\rangle|\downarrow\rangle \\ +\textstyle\sqrt{\frac{2}{3}} |\sin\theta \cos\phi \hat{x} + \sin\theta \sin\phi \hat{y}+\cos\theta \hat{z}\rangle|\uparrow\rangle,
\end{split}
\end{equation}
where $\theta$ is the angle between the normal to the sample and the polarization vector of the incoming light, and $\phi$ is the angle of rotation about the normal.

Calculating the optical momentum matrix element between these states and the conduction band states for the ``dark" exciton, we obtain
\begin{equation}
\begin{split}
\textstyle \langle iS\uparrow |{\bf p} |\frac{3}{2},\frac{1}{2}\rangle = \textstyle\sqrt{\frac{2}{3}} (\sin\theta \cos\phi \hat{x} + \sin\theta \sin\phi \hat{y}+\cos\theta \hat{z})P
\end{split}
\end{equation}
\begin{equation}
\begin{split}
\textstyle \langle iS\downarrow |{\bf p} |\frac{3}{2},-\frac{1}{2}\rangle  = \textstyle\sqrt{\frac{2}{3}} (\sin\theta \cos\phi \hat{x} + \sin\theta \sin\phi \hat{y}+\cos\theta \hat{z})P,
\end{split}
\end{equation}
where ${\bf p}$ is the dipole operator, $|S\uparrow\rangle$ and $|S\downarrow\rangle$ are the conduction band eigenstates, and $P$ is the dipole element.

If the plane of incidence is the $x-z$ plane, then $\phi=0$ and the TE polarization is along the $y$-axis. We obtain the relevant matrix elements by looking at the $\hat{y}$ component which gives us $\sin\theta \sin\phi = 0$ since $\phi=0$. To obtain the matrix element corresponding to the TM polarization, we look at the $\hat{z}$ component which gives us $\cos\theta$. In our experiment, we measure the angle $\theta'$ where $\theta'$ is the angle between the normal to the sample and the incident beam and $\theta'=\frac{\pi}{2}-\theta$, giving us a matrix element $\sim\sin\theta'$ $\sim k_x$.

The dependence of the S term on $k_x$ and the dependence of the polarization selection rule on $k_x$ gives us a factor of $k_x^2$ for the matrix element. The rate of two-photon absorption is proportional to the square of the matrix element, which means that we should see the intensity increase as $k_{||}^4$ for TM polarized light, and we should not see any luminescence from a purely TE polarized light. To check this, we varied the k for incident TM and TE polarized light, and observed the intensity consistent with a $k_{||}^4$ dependence in the TM polarized case (Figure 6), and no luminescence in the TE polarized case. The spread of angles in the incoming beam gives a non-zero contribution even at $k_{||}=0$, i.e., normal incidence. When the signal was sent through a polarizer, the polaritons formed are seen to be TM polarized as well. A change of the polarization of the pump beam from linear to circular polarization causes a decrease in absorption. 

\begin{figure}
\centerline{\includegraphics[width=0.8\columnwidth]{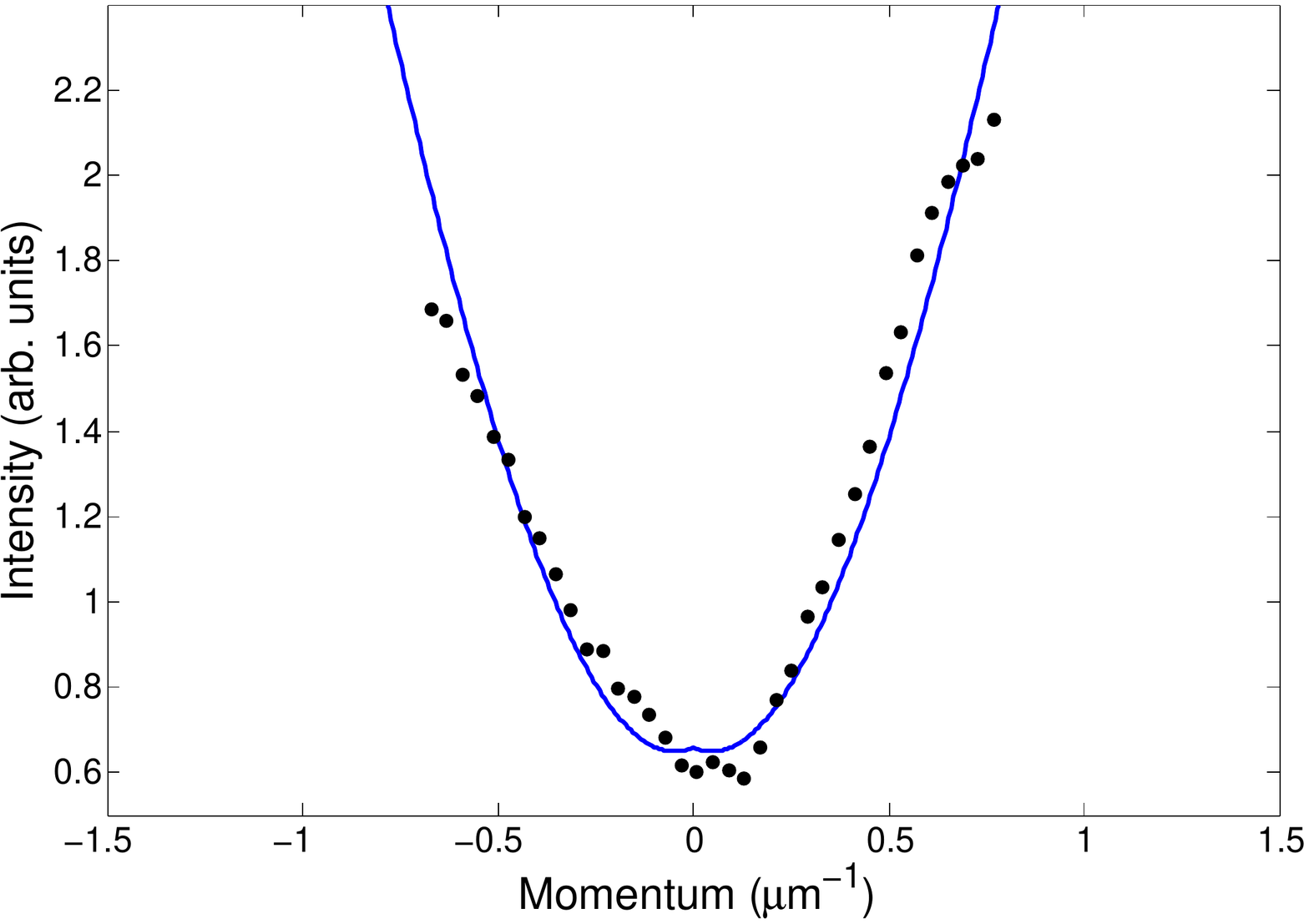}}
\caption{Intensity of two-photon absorption vs. the central in-plane momentum.]{Intensity of the light from two-photon absorption vs. the central in-plane momentum of the incoming light beam. The dots represent our data and the solid line is a convolution fit $A(f*g)+B$, where $f=k^4$ and $g = e^{-\frac{k^2}{\sigma^2}}$ with $\sigma^2=0.07$, which takes into account the finite width of our laser spot. $A=0.95$ and $B=0.14$ are scaling parameters which take into account background light. This indicates that the absorption is $\propto k^4$.}}
\end{figure}

Further support of this conclusion comes from experiments in which we placed the sample in a magnetic field. No change in the intensity of the polariton emission generated by two-photon excitation was seen. Since the ``dark" state/``bright" state mixing is already allowed at finite in-plane $k_{||}$, a change in the magnetic field up to $\sim 1$~T did not give a substantial increase in the mixing.

In conclusion, direct two-photon excitation of the lower polariton branch of exciton-polaritons in a microcavity is possible at non-zero angle of incidence, without involving higher-lying $J=2$ or $2p$ exciton states. When the pump photon energy is tuned to be resonant with those higher-lying states, we do see evidence for those states being excited, which then lead to polaritons appearing at lower energy with a long rise time. Direct two-photon excitation of polaritons leads us to expect novel nonlinear effects with interaction of macroscopically occupied polariton states and light waves at half their frequency. Future work will address this.

The work at the University of Pittsburgh was supported by the National Science Foundation under grants PHY-1205762. The work at Princeton University was partially funded by the Gordon and Betty Moore Foundation as well as the National Science Foundation MRSEC Program through the Princeton Center for Complex Materials (DMR-0819860).



\end{document}